\documentclass[psfig]{elsart}
\usepackage{amssymb}
\usepackage{alltt}

\begin{document}
\begin{frontmatter}
\title{Logic for Unambiguous Context-Free Languages}
\author{Yassine Hacha\"{\i}chi}
\address{LAMSIN - ENIT, Universit\'e de Tunis El Manar}
\ead{hachaichi.ens@gmail.com}
\begin{abstract}
We give in this paper a logical characterization for unambiguous Context Free 
Languages, in the vein of descriptive complexity. A fragment of the logic 
characterizing context free languages given by Lautemann, Schwentick and Th\'erien
 based on implicit definability is used for this aim. We obtain a new connection
  between two undecidable problems, a logical one and a language theoretical one.
\end{abstract}
\end{frontmatter}
\section{Introduction}
A language $L$ over an alphabet $A$ can be defined by several manners. The most famous are:
\begin{enumerate}
\item a subset of $A^*$ whose elements satisfy some given property; it is the analogous of comprehension schema in set theory,
\item a subset of $A^*$ whose elements are generated by some formal grammar,
\item a subset of $A^*$ whose elements are recognized by some model of computation.
\end{enumerate}
In complexity theory, we try to classify languages according to the recognizer used (finite automata, push down automata, $\cdots$),
or by the ressources (time, space, $\cdots$) neded by some model of computation (turing machine, random access machine, $\cdots$), see \cite{PAP}
for a detailed introduction to the field.

One of the aims of \emph{descriptive complexity} \cite{EF,Str} is to evaluate how easy or hard it is to express a given property defining some language (as in 1 above) in the language of logic.

The answer to this question got a meaning by the works of B\"uchi \cite{buc} and Elgot \cite{elg} who made the link between formal logic and formal language theory. This connection was made by identifying words to {\it finite} logical structures. Their result was that a word language is regular if, and only if it is the class of models of some Monadic Second-order sentence. Two questions were naturally asked:

The first one is:
\begin{enumerate}
\item What is the expressive power of Monadic Second-order Logic on other structures than words, graphs and trees for example?

The other question is:

\item Is there a logical description for each known class of words: star free, context free, $\ldots$?
\end{enumerate}

Both directions were explored since, we will recall some results in the next section.

The logical description of the behaviour of computational models was also taken up in complexity theory. Starting with Fagin's work, it was shown that many complexity classes such as NP, P, LogSpace, NLogSpace, Pspace, $\cdots$ could be characterized by different varieties of second-order logic (involving for example fixed point logic or transitive closure operator). For an introduction to this field see Ebbinghaus and Flum's book \cite{EF}.

In \cite{HY1} and \cite{HY2}, I used some generalized quantifiers of comparison
of cardinality, to get a new logical characterizations of the class of rudimentary languages in
the scope of descriptive complexity.
Lautemann, Schwentick and Th\'erien \cite{LST} gave recently a logical description of Context Free Languages.
They used for this purpose the semantic quantifier of matching.

Our contribution in this paper is, in a first time,
using a result of McNaughton and Papert \cite{MP} we will refine an algebraic normal form of Chomsky and Sch\"utzenberger which characterizes Context Free Languages using the Dyck languages.

The Second result of this paper is a description of Unambiguous Context Free Languages. This description uses a fragment of a logic built from first-order implicitely definable predicates introduced by Kolaitis \cite{Kol}. This logic was motivated by the failure of the Beth property when we confine ourselves to finite structures.

This paper is organized as follows:

In the next section we give some background of language theory and logic, and we introduce some results of descriptive recognizability. We introduce in the last subsection the result of Lautemann and {\it al} \cite{LST}.

In section 3 we refine an algebraic normal form given by Chomsky and Sch\"utzenberger \cite{CS}, for describing Context Free Languages using the Dyck language.

In section 4 we give the logical characterization of unambiguous context free languages.

In the conclusion we try to link undecidability of unambiguity and undecidability of the logic $IMP$.

\section{Notations and Background}

We give here some definitions and results in formal language theory, logic and the connection between them.

For the rest of the section $\Sigma$ will denote a finite vocabulary $\{ c_1,\ldots ,c_s\}$. A language is a subset of $\Sigma^*$, which is the set of finite words on $\Sigma$.

\subsection{Formal Language Theory}

We will recall in this section some notions of language theory, from the grammatical point of view, which we will use later in this paper, the curious reader can find more details on this area in Harrison's book \cite{Ha}.

A {\it context free grammar} is a 4-tuple $ < \Sigma , N , S , P >$ such that:
\begin{itemize}
\item $\Sigma$ and $N$ are finite disjoint sets, called respectively the set of {\it terminal} and {\it non-terminal} symbols,
\item $S$ is a special symbol of $N$, called the {\it start symbol} or the {\it axiom} of the grammar,
\item $P$ is a {\it set of productions} of the form $X \rightarrow w$, where X is a non-terminal and $w \in (\Sigma \cup N)^*$.
\end{itemize}

If we replace each non-terminal symbol by a new symbol $|$ not in $\Sigma \cup N$ in the right-hand side of a production we obtain a string called the {\it pattern} of the production.

A context free grammar is {\it regular} if all productions are of the form $X \rightarrow w|w Y$ where X and Y are non-terminals and $w \in \Sigma^*$.

We define the (one step) {\it derivation rule} $\Rightarrow _G$ for a grammar $G$ by 

\begin{center}
$ w_1Xw_2 \Rightarrow _G w_1ww_2 $ is a derivation if and only if $ X\rightarrow w \in P$.
\end{center}

The reflexive and transitive closure of $\Rightarrow _G$ is denoted ${\stackrel{*}{\Rightarrow}}_G$.

A language $L \subseteq \Sigma^*$ is {\it context free} (resp. {\it regular}) if and only if there is a context free (resp. regular) grammar which derives it from $S$, i.e \\$L=L(G)=\{w \in \Sigma | S {\stackrel{*}{\Rightarrow}}_G w \}$.

A language is {\it star free} if it is build from finite languages by only boolean operations and concatenation.

The {\it derivation tree} of a word $w$ associates naturely to the derivations made from $S$ until reaching $w$.

Formally, a {\it derivation tree of a word} $w \in L(G)$ is a tree so that :

\begin{itemize}
\item the root is labelled by the start symbol S,
\item the leaves are labelled by terminals,
\item the internal nodes are labelled by non terminals,
\item the passage from an internal node to its sons corresponds to a production,
\item the lecture of leaves from left to right give $w$.
\end{itemize}

{\bf Example} Let's take the grammar 
$$G = <\{a,b\},\{X_0,X_1,X_2,X_3,X_4\},X_0,P>$$
 where $P$ contains the following productions:

\begin{tabular}{l}
$P_{0,1}:X_0 \rightarrow aX_1X_2ba$ \\
$P_{1,1}:X_1 \rightarrow aX_3X_2b$\\
$P_{2,1}:X_2 \rightarrow aab$\\
$P_{2,2}:X_2 \rightarrow ab$\\
$P_{3,1}:X_3 \rightarrow ab$\\
\end{tabular}

Let $w=aaababbaabba$ the word with derivation tree given in figure 1.
\begin{figure}[h]
\begin{center}
\setlength{\unitlength}{1cm}
\begin{picture}(5,4)(0,0)
\put(0,0){\includegraphics{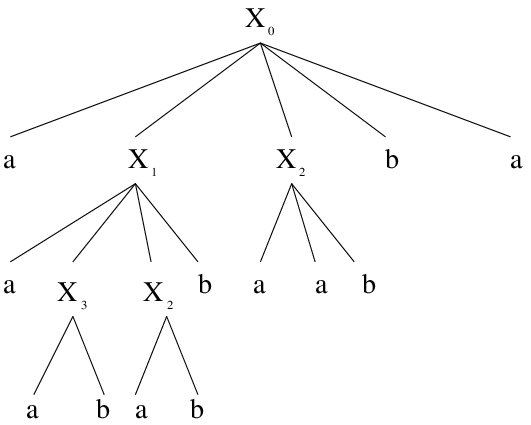}}
\end{picture}
\caption{A derivation tree of $w$}
\end{center}
\end{figure}

A context free grammar is {\it unambiguous} if every word in $L(G)$ has a unique derivation tree.

A context free language is {\it unambiguous} if it has an unambiguous context free grammar which derives it.

\subsection{Logic}
As we mentioned in the introduction, we can identify words to finite models in a special logical signature $\tau_\Sigma = \{ < , P_{c_1},\ldots ,P_{c_s}\}$.

In this introductory section to logic we define formally the word model and some logics that we will use in the forthcoming sections.

We will confine ourselves to finite structures and especially to words.

Let $\Sigma = \{ c_1,\ldots ,c_s\}$ be a finite vocabulary. We associate with a word 
$w=a_1 \ldots a_n$ over $\Sigma$, the {\it word model} $S_w$, namely the relational structure $S_w=(\{1,...,n\},<, P_{c_1},\ldots ,P_{c_s})$, where $<$ is the natural order on $\{1,...,n\}$ and $P_a$ is 
the unary predicate collecting the positions of $w$ labelled $a$:\\
$P_a = \{ i \in N | a_i =a \}$.

For example let's take the word $w=aabbabb$ on the vocabulary $\Sigma=\{ a, b\}$.

The corresponding logical structure will be:

$S_w = ( \{1,2,3,4,5,6,7\},<,P_a^w = \{1,2,5\},P_b^w =\{3,4,6,7\})$.

The set of {\it first-order formulas} (on words), F.O., is built inductively from atomic formulas of the form:

\begin{center}
$x<y$ and $P_a(x)$ for $a \in \Sigma$.
\end{center}

by means of connectives $\land , \, \lor , \, \rightarrow , \, \leftrightarrow ,\, and \; \neg$ and the quantifiers $\exists \; and \; \forall$.

Formally:
\begin{enumerate}
\item If $x, \; y$ are any variable symbol or constant symbol then $x<y$ and $P_a(x)$ for $a \in \Sigma$ are first-order formulas.
\item If $\phi, \; \psi$ are first-order formulas then 
$$\phi \land \psi , \; \phi \lor \psi , \; \phi \rightarrow \psi, \; \phi \leftrightarrow \psi, \; and \neg \phi$$
 are also first-order formulas.
\item If $\phi$ is a first-order formula and $x$ is a variable then $\exists x \phi,$ and $\forall x \phi$ are first-order formulas.
\end{enumerate}

{\it Monadic second-order logic}, M.S.O., is built like first-order logic where we add $X_i(x)$ to atomic formulas (the first item in the first-order construction), for some unary variables $(X_i)_{i \in N}$ and can quantify over $X_i's$ (in the last item of the first-order construction).

A language L is said to be (explicitly) {\it definable in a logic} $\lambda$ if and only if there exists a formula $\phi \in \lambda$ such that:

\begin{center}
$\forall w \in A^* ( w \in L \Leftrightarrow S_w \models \phi)$.
\end{center}

Most classical model theoretical results fail when we confine ourselves to only finite models \cite{EF}, especially Beth's theorem.

Let R be an n-ary relation symbol not in $\tau_\Sigma$. An F.O.
$[\tau_\Sigma\cup \{R\}]$ sentence,
{\it $\phi$, defines R implicitly} if every word structure has at most one expansion $(S_w,R)$ to a 
$\tau_\Sigma\cup \{R\}$-structure satisfying $\phi$.

The Beth's theorem says that a predicate is implicitly definable in F.O. if and only if it is explicitely definable in F.O., in the class of all structures (finite and infinite).

The failure of the Beth property for finite models, (see \cite{EF} for all properties of (classical) model theory that fail in finite model theory), stimulated Kolaitis \cite{Kol} to define a logic of implicitly defined queries in some logic.

We will only use the case of first-order implicitely defined queries.

\textsc{Remark.}
We use a different definition from the one used by Kolaitis in \cite{Kol}. For a discussion see \cite[page 213]{EF}

The logic $IMP=IMP(F.O.)$ is the set of formulas, that allows to define exactly those queries (properties) that are expressible in F.O. using first-order implicitely definable queries.

Formally:

An {\it $IMP$-formula} $\phi({\overline x})$ is a tuple, 
$$(\psi_1(R_1),\ldots,\psi_m(R_m),\psi({\overline x},R_1,\ldots,R_m))$$
where all $\psi_i's$ are first-order sentences on $\tau_\Sigma \cup \{ R_i \}$, and $\psi$ 
 is first-order on $\tau_\Sigma \cup \{ R_1, \ldots ,R_m \}$
\\such that $\models_{fin} \exists! R_1 \ldots \exists! R_m (\psi_1(R_1) \land \ldots \land \psi_m(R_m))$.\\
Where $\models_{fin}$ means satisfiability on finite models, and $\exists !$ means there is a unique.

The meaning of $\phi({\overline x})$ is fixed by requiring that:
$$S_w \models \forall X_1 \ldots \forall X_m (\psi_1(X_1) \land \ldots \land \psi_m(X_m))$$
$$ \rightarrow \forall {\overline x} (\phi({\overline x}) \leftrightarrow \psi({\overline x},X_1, \ldots ,X_m))$$
This means: $S_w \models \phi({\overline a}) \Leftrightarrow S_w \models \psi({\overline a},R_1,\ldots,R_m)) $, where the $R_i's$ are uniquely determined by
 $S_w \models \psi_i(R_i)$.

If we use only unary predicates we will denote this logic {\it $IMP(1)$}.

\subsection{Logic vs. Language theory}

We will give in this section some results, in chronological order of their publications, on logical characterization of classes of languages, and the link between logical complexity and recognizability complexity.

\begin{thm}[B\"uchi \cite{buc}, Elgot \cite{elg} and Lacroix \cite{LA}]
A language is regular if and only if it is definable in monadic second-order logic if and only if it is definable in IMP(1).
\end{thm}

Thomas improved the result of B\"uchi and Elgot to formulas of the form $\exists P \, F.O.$ where P is a monadic predicate.

The connection between trees and words was given by Mezei and Wright, for a comprehensive proof and a definition of regular tree languages see \cite{GS}.

\begin{thm}[Mezei and Wright \cite{MW}]
A word language is context free if and only if the set of the derivation trees of its words is regular.
\end{thm}

And a new description of regular tree languages in the same vein as B\"uchi's result arises,

\begin{thm}[Doner, Thatcher and Wright \cite{DD,TW}]
A tree language is regular if and only if it is definable in monadic second-order logic.
\end{thm}

First-order logic, the most natural sublogic of monadic second-order logic, on words was studied by McNaughton, for more details see \cite{Str}.

\begin{thm}[McNaughton and Papert \cite{MP}]
A language is star free if and only if it is definable in first-order logic.
\end{thm}

Fagin studied the Existensial fragment of Second-order Logic and proved:

\begin{thm}[Fagin \cite{fagin}] Languages definable in existential second-order logic are exactly those computable in polynomial time by a non deterministic Turing machine.
\end{thm}

This result is true for all finite relational structures not only for word structures.

\begin{thm}[Thomas, Perrin and Pin \cite{Th}] A star free set is of dot depth $n$ 
if and only if it is definable in the boolean closure of $\Sigma_n$, Where $\Sigma_n$ is the set of first-order formulas allowing n alternations of quantifiers (universal, existential).
\end{thm}

A recent result of Eiter and {\it al} \cite{GGG}, which says that every Existential Second-order prefix class either describes only regular languages or describes an NP-complete problem.

I suggest to curious readers to see the expository papers of Thomas \cite{Th}, or Pin's one \cite{Pi} for more details and results.

\subsection{Matching vs. Context free languages}

In this section we recall a result of Lautemann, Shwentick and Th\'erien for the description of Context Free Languages using the semantic quantifier of Matching.

\begin{defn}
A binary relation $M$ is called a {\it matching} if it satisfies the following conditions :
\begin{enumerate}
\item  $\forall ij[(i,j) \in M \Rightarrow i<j]$ .
\item  $\forall ij[(i,j) \in M \Rightarrow \forall k \neq i,j$\\ $((i,k),\, (k,i),\, (j,k),\, and (k,j) \;are\; not\; in\; M)]$.
\item  $\forall ijkl[(i,j),(k,l) \in M \Rightarrow (i<k<j \rightarrow i<l<j)]$.
\end{enumerate}

We will denote by $\psi (M)$, the conjunction of these three first-order items on $\tau \cup \{M\}$.

Let {\it Match} denote the class of matchings on word structures.
\end{defn}

$\exists Match \,\phi$ means : there exists a relation $M \in Match$ such that $<S,M> \models \phi$.

In order to define any non regular language, one have to go beyond M.S.O. Logic. On the other hand, existential quantification over a simple binary relation 
express all context free languages and some $NP-$complete languages by the result of Eiter and {\it al} \cite{GGG}, their result says that any prefix class of Existential Second-order Logic either expresses only regular languages or expresses some $NP-$complete language.

Moreover they proved that $NP$-hardness is present with a sentence 
$\exists R \phi$ where $R$ is a binary predicate and $\phi$ is first-order of the appropriate form.

Lautemann and {\it al} \cite{LST} choose a semantical approach for this purpose, in which they restrict the second-order quantifier to range over the class of matchings, {\it Match}. 
They define the class $\exists Match \,F.O.$ to consist of all those sets $L$ of $\tau-$structures for which there is a first-order sentence $\phi \; over \; \tau \cup \{M\}$ such that, For every 
$\tau-$structures $S_w$:\\
$S_w \in L$ if and only if, there is some matching M over $S_w$ such that 
$<S_w,M> \models \phi$.

They proved the following:

\begin{thm}[Lautemann, Schwentick and Th\'erien \cite{LST}]
A language L is context free if and only if it is definable by a formula of the form $\exists Match \,\phi$ where $\phi \in F.O.$

The result remains true also for $\phi \in M.S.O.$
\end{thm}

For proving this result, the first step was to construct a first-order sentence over $\tau \cup \{M\}$ 
for each grammar G, which holds for a word structure $S_w$ if and only if there is a G-derivation tree T of $w$, 
and there is an effective way to construct the matching from the tree, and vice versa.

For the other direction they combined results of Doner \cite{DD}, Thatcher and Wright \cite{TW} and Mezei and Wright \cite{MW} to have the fact: 
\emph{A language is context free if and only if the set of derivation trees of his words is a regular tree language.} (And this is independent of the grammar used.)

The last step was to construct trees from the matching, and prove that these trees satisfy some monadic second-order sentence. 
More details will be given in the proof of the theorem 10.

\section{A new Chomsky and Sch\"utzenberger Normal form}
In this section we reprove a stronger version of the Chomsky and Sch\"utzenberger theorem, by restricting the expression K, of theorem 8, to be only star free. This result is also given in \cite{aut}, but we reprove it by only logical arguments.

\begin{thm}[Chomsky and Sch\"utzenberger \cite{CS}]
A language L in $\Sigma^*$ is context free if and only if  $L=\psi(D_n \cap K)$ where $D_n$ is the Dyck language on n ``brackets'', K a regular expression and $\psi$ a mono\"{\i}d homomorphism from  $\Gamma \cup \overline{\Gamma} \, into \, \Sigma^*$.
\end{thm}

We recall the Dyck language $D_n$ on n brackets is the language generated by the grammar:

$G=<\{a_1,\ldots ,a_n,{\overline a_1},\ldots ,{\overline a_1}\},\{S\},S,P>$ where P is the set of productions:

$S \rightarrow a_1S{\overline a_1}S | \ldots |a_nS{\overline a_n}S|\varepsilon$ where $\varepsilon$ denotes the empty word.

If the $a_i's$ are assumed to be the opening brackets and ${\overline a_i}'s$ the closing ones, $D_n$ will be the set of well balanced brackets words.

\begin{thm}
A language L is context free if and only if  $L=\psi(D_n \cap K)$ where $D_n$ is the Dyck language on n ``brackets'', K a star-free expression and $\psi$ a mono\"{\i}d homomorphism from  $\Gamma \cup \overline{\Gamma} \, into \, \Sigma^*$.
\end{thm}
We recall the double Greibach normal form.

\begin{lem} Every context free language is generated by a grammar $G=<N,\Sigma,S,P>$ which satisfies the following condition:\\
 all productions are of one of the forms:
\begin{enumerate}
\item $S \rightarrow a ,\, a \in \Sigma$ or 
\item $X \rightarrow a u b , \, X \in N , \, a,b \in \Sigma , \, and \, u \in (\Sigma \cup N )^*.$
\end{enumerate}
\end{lem}

For a detailed proof we send the reader to the paper of Autebert and {\it al} \cite{ABB}.

{\bf Proof of the Theorem} The way $L=\psi(D^*_n \cap K)$ for some star free expression K implies that L is a context free language derives obviously from the Chomsky and Sch\"utzenberger theorem because star free expressions are regular.

For the other way we will give some first-order conditions on a Dyck language to construct a set Z such that $L=\psi(Z)$. These conditions are intimately connected to the history of derivations.

Let L be a context free language. By the previous lemma we have a grammar in double Greibach normal form $G=<\Sigma,N,S,P>$ which derive it from S.

We enumerate first the non-terminal symbols, $X_0=S , ... , X_N$.

After we label productions by ordered pairs $<i,j>$ where $X_i$ is the left hand side non terminal of the production, and $j$ enumerates injectively the productions having $X_i$ as left hand side non terminal. The elements of P are:

\begin{tabular}{l}
$P_{0,1}:X_0 \rightarrow w_{0,1}$ \\
~~~~~$\ldots$ \\
$P_{0,i_0}:X_0 \rightarrow w_{0,i_0}$\\
~~~~~$\ldots$\\
$P_{N,1}:X_N \rightarrow w_{N,1}$\\
~~~~~$\ldots$ \\
$P_{N,i_N}:X_N \rightarrow w_{N,i_N}$\\
\end{tabular}

and for each production we denote $c_{i,j}$ the total 
number of right hand side non terminals in the production $p_{i,j}$.

We construct now the set of brackets $\Gamma$. It is the set of tuples of integers $<a,b,c,d,e,f>$ where:

\begin{description}
\item[a, b] are a production code such that $c_{a,b} \neq 0$ or $a=b=0$.
\item[c] is 1 if $a=b=0$, else $c=c_{a,b}$.
\item[d] is such that $1 \leq d \leq c$ and represents the range of the current non terminal in $p_{a,b}, \; 1\leq d \leq c$.
\item[e, f] are the next production code where e must be the code of the $c^{th}$ non terminal in the right hand side of the production $p_{a,b}$ and $f \leq i_e$ ,or $e=0$ if $a=b=0$.
\end{description}

$\overline{\Gamma}$ will be the set of ${\overline {<a,b,c,d,e,f>}}$ for each element \\ $<a,b,c,d,e,f> \in \Gamma$.

We give now the conditions on the words of the Dyck language on $D_\Gamma$ to be in Z.

We will decide of the successor of each symbol in this word and give the range of the first and the last symbol.

\begin{enumerate}
\item The first symbol in our word must be an opening bracket of a start configuration and the last one must close this bracket $ \bigvee_{1 \leq i \leq i_0}( P_{<0,0,1,1,0,i>}(min) \land P_{\overline {<0,0,1,1,0,i>}}(max))$.
\item $P_{<a,b,c,d,e,f>}(x)$ and $c_{e,f}=0$ then we must close immediately our bracket $P_{\overline {<a,b,c,d,e,f>}}(x+1)$ because $p_{e,f}$ is a terminal production.
\item $P_{<a,b,c,d,e,f>}(x)$ and $c_{e,f}>0$ then we have\\ $\bigvee_{e',f'} P_{<e,f,c_{e,f},1,e',f'>}(x+1)$ such that $e'$ is the first non terminal in the right hand side of $p_{e,f}$.
\item $P_{\overline {<a,b,c,d,e,f>}}(x)$ for some x and $c>d$ then we must have $\bigvee_{e',f'} P_{<a,b,c,d+1,e',f'>}(x+1)$ for some $f'<i_e'$ such that $e'$ is the $d+1^{st}$ non terminal in $p_{a,b}$.
\item $P_{\overline {<a,b,c,d,e,f>}}(x)$ for some x and $c=d$ then we must have $\bigvee_{<a,b,c,d,e,f>} P_{\overline {<a,b,c,d,e,f>}}(x+1)$.
\end{enumerate}
We are sure in item 5 to close the good type of parenthese because we are in a Dyck language.

Because of the finiteness of the set $\Gamma$ these conditions are expressed by a first-order formula.

By McNaughton and Papert's theorem Z is a star free subset of $D_\Gamma$.

If $c_{a,b} \neq 0$ then the production $p_{a,b}$ have the form:

$p_{a,b}:X_a \rightarrow w_{(a,b,0)}X_{j_1} \ldots w_{(a,b,c_{a,b}-1)}X_{j_{c_{a,b}}} w_{(a,b,c_{a,b})}$.

We now give the homorphism $\phi$:

$\phi (<a,b,c,d,e,f>) = w_{(e,f,0)} $, and\\
$\phi({\overline {<a,b,c,d,e,f>}})=w_{(a,b,d)}$, and\\
$\phi({\overline {<0,0,1,1,0,i>}})=\varepsilon$.

Where $\varepsilon$ is the empty string.

By identifying the brackets to internal nodes of the spanning tree and the homomorphism images to leaves in the right place, we can trivially verify the eqality  $L=\psi(Z)$. {\bf Q.E.D}

{\bf Example} Let's take the grammar 
$$G = <\{a,b\},\{S,Y,Z \},S,P>$$ 
where P contains the following productions:

\begin{tabular}{l}\\
$S \rightarrow abba|aYabZba$ \\
$Y \rightarrow aaYbaZbb|aZb$\\
$Z \rightarrow ab$\\
\end{tabular}\\

We first enumerate the non-terminals: $S=X_0,\; Y=X_1,\; and \; Z=X_2$. We can now enumerate productions:

\begin{tabular}{l}\\
$p_{0,1}: X_0 \rightarrow abba$ \\
$p_{0,2}: X_0 \rightarrow aX_1abX_2ba$\\
$p_{1,1}: X_1 \rightarrow aaX_1baX_2bb$\\
$p_{1,2}: X_1 \rightarrow aX_2b$\\
$p_{2,1}: X_2 \rightarrow ab$\\
\end{tabular}\\

So we have: \\
$\Sigma=
\{ \langle 001101\rangle ,\langle 001102\rangle ,\langle 022111\rangle ,\langle 022112\rangle ,\langle 022221\rangle ,$\\
$\langle 112111\rangle ,\langle 112112\rangle ,\langle 112221\rangle ,\langle 121121\rangle \}$

Which we will denote later 1, 2, 3, 4, 5, 6, 7, 8, and 9.

The Dyck words must satisfy the formula

$$\begin{array}{ll}
F  \equiv&   (((P_1(min)\wedge P_{\overline{1}}(max))\lor (P_2(min)\wedge P_{\overline{2}}(max)))  \wedge \\
&(P_1(x) \rightarrow P_{\overline{1}}(x+1)) \wedge \\
&(P_2(x) \rightarrow (P_3(x+1) \lor P_4(x+1))) \wedge \\
&(P_3(x) \rightarrow (P_6(x+1) \lor P_7(x+1))) \wedge \\
&(P_4(x) \rightarrow P_9(x+1)) \wedge \\
&(P_5(x) \rightarrow (P_{\overline{5}}(x+1)) \wedge \\
&(P_6(x) \rightarrow (P_6(x+1) \lor P_7(x+1))) \wedge \\
&(P_7(x) \rightarrow P_9(x+1)) \wedge \\
&(P_8(x) \rightarrow (P_{\overline{8}}(x+1)) \wedge \\
&(P_9(x) \rightarrow (P_{\overline {9}}(x+1)) \wedge \\
&(P_{\overline{1}}(x) \rightarrow x=max) \wedge \\
&(P_{\overline{2}}(x) \rightarrow x=max) \wedge \\
&(P_{\overline{3}}(x) \rightarrow (P_5(x+1) ) \wedge\\ 
&(P_{\overline{4}}(x) \rightarrow (P_5(x+1) ) \wedge \\
&(P_{\overline{5}}(x) \rightarrow \bigvee\limits_{1 \leq i \leq 11} P_{\overline{i}}(x+1)) \wedge \\
&(P_{\overline{6}}(x) \rightarrow (P_8(x+1) ) \wedge \\
&(P_{\overline{7}}(x) \rightarrow (P_8(x+1) ) \wedge \\
&(P_{\overline{8}}(x) \rightarrow \bigvee\limits_{1 \leq i \leq 11} P_{\overline{i}}(x+1)) \wedge \\
&(P_{\overline{9}}(x) \rightarrow \bigvee\limits_{1 \leq i \leq 11} P_{\overline{i}}(x+1)).
\end{array} $$
The homomorphism $\phi$ is defined by:

$$\begin{array}{lcl}
\phi (1)= abba& , &\phi (2)=a\\
\phi (3)=aa& , &\phi (4)=a\\ 
\phi (5)=ab& , &\phi (6)=aa\\
\phi (7)=a& , &\phi (8)=ab\\
\phi (9)=ab& , &\phi({\overline {1}})=\varepsilon\\
\phi({\overline {2}})=\varepsilon& ,&\phi({\overline {3}}) = ab\\ 
\phi({\overline {4}}) = ab&,& \phi({\overline {5}}) = ba\\ 
\phi({\overline {6}}) = ba&,& \phi({\overline {7}}) = ba\\ 
\phi({\overline {8}}) = bb& and &\phi({\overline {9}}) = b\\
\end{array} $$

Let's take as example the word $w=aaabbababba$, we give in the figure below its derivation tree.
\begin{figure}[t]
\begin{center}
\setlength{\unitlength}{1cm}
\begin{picture}(5,6)(0,0)
\put(0,0){\includegraphics{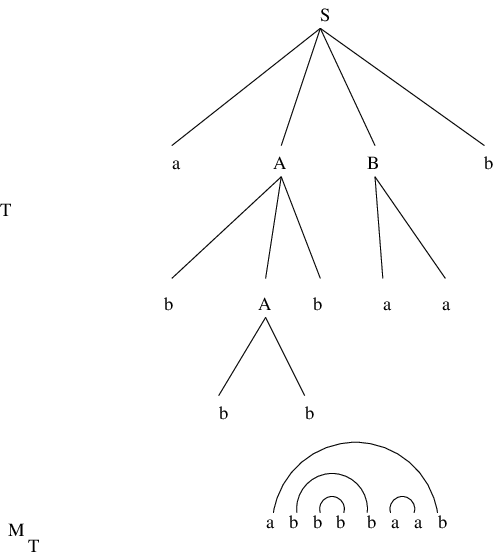}}
\end{picture}
\caption{Matching from derivation tree of $w$}
\end{center}
\end{figure}

Then by extracting in a prefixed (first reach) way the opening brackets and at the same time in a postfixed (last reach) way the closing ones we get the word $w_D \in D^*$ \\
\begin{center}
$w_D = 2 \; 4 \; 9 \; {\overline 9} \; {\overline 4} \; 5 \; {\overline 5} \; {\overline 2}$
\end{center}

We then remark that $\phi (w_D) = w$.

The construction is closely connected to the derivation tree this is why we are sure of the equivalence.

\section{A logic for unambiguous Context free languages}

We give now a logic for unambiguous Context free languages.
The main idea is that unambiguity needs unicity of existence.

Let the Logic $IMP_2$ be the sublogic of $IMP$ where we use the implicit definition of only one predicate, which is binary.

$\exists Match \; F.O. \cap IMP_2$ will be the set of formulas of $\exists ! Match \; F.O.$ where only one matching M satisfy the first-order formula.
\begin{thm}
A language is unambiguous context free if and only if it is definable in\\$\exists Match \; F.O. \cap IMP_2$.
\end{thm}

One of the keysteps in the proof is:

\begin{lem}[\cite{GH}]
Every Unambiguous Context Free Language has an Unambiguous Context Free Grammar $G=<N,\Sigma,S,P>$ where all productions are of one of the forms:
\begin{enumerate}
\item $S \rightarrow a ,\; a \in \Sigma $ or 
\item $X \rightarrow a u b , \, X \in N , \, a,b \in \Sigma , \, and \, u \in (\Sigma \cup N )^*.$
\end{enumerate}
\end{lem}

This lemma uses only the fact that the classical construction preserves unambiguity.

{\bf Proof of the theorem.} The proof of this theorem is intimately connected to the one of Lautemann and {\it al} for giving a logic for context free languages, we only have to prove that unambiguity of the language implies uniqueness of the matching and vice versa.

By the previous lemma we have an unambiguous grammar in the normal form used in \cite{LST}. The processes:
\begin{enumerate}
\item Eliminate all productions of the form $X\rightarrow \alpha$ for some $\alpha \in \Sigma$ by introducing a new production $Y \rightarrow u\alpha v$, for every production $Y \rightarrow u X v \in P$.
\item Enumerate all non-terminals, $X_1=S , ... , X_N$. Starting with $i=2$ do the following for every $i$, as long as there is non-terminal production p= $X_i \rightarrow v$ whose pattern also appears as the pattern of a production with left-hand side $X_j,\; j<i$ replace p by all productions which can be obtained from it by substituting one of the non-terminals in $v$ in all possible ways.
\end{enumerate}
terminates and preserves unambiguity.

Then for every Unambiguous Context Free Language we have a Unambiguous Context Free Grammar in double Greibach Normal Form and any two non terminal productions have the same pattern iff they have the same left hand non terminal.

Let $T$ be a derivation tree of $w$, the matching corresponding to T is $M_T$ defined by:$(i,j) \in M_T$ if and only if i corresponds to the leftmost and j to the rightmost child of the same internal node of T.

We construct now the formula $\psi_G$ over $< \Sigma , < , M>$ which holds for a
string $w$ with matching M iff there is a G derivation tree T for $w$ such that $M=M_T$. 
It follows that there is a matching M on $w$ with 
$<w,M> \models \psi_G$ iff $w$ can be derived in G.

Let $(i,j) \in M_T$ an arch, the pattern of $(i,j)$ is the string composed of
their ``brothers'' written from left to right where internal nodes are replaced by$|$.

To be the matching constructed from a G derivation tree, the pattern must
correspond to the pattern of a production in G.

For $p \equiv X_0 \rightarrow \alpha v_0 X_1v_1 \ldots X_s v_s \beta \; where \;
\alpha,\, \beta \in \Sigma,\; v_i \in \Sigma^*$, and 
$X_i \in N$ we construct a first-order formula:
$\pi_p(x,y)=$\\
$P_\alpha(x) \land P_\beta(y) \land \exists x_1 y_1 \ldots x_s y_s
[(x<x_1<y_1< \ldots < x_s<y_s<y)$\\
$\land 
(\psi_{v_0}(x,x_1)\land \psi_{v_1}(y_1,x_2)\land \ldots \land \psi_{v_s}(y_s,y))$\\
$ \land
(M(x_1,y_1) \land \ldots \land M(x_s,y_s))]$,

where $\psi_{v}(i,j)$ is the first-order formula
$\bigwedge ^{n=j}_{n=i} P_{w_{n-i}}(n)$ if $v= w_0 \ldots w_r$, Which characterize the pattern between two
positions $x \; and \; y$ to correspond to some production p of G.

Let $\pi_X(x,y)$, for $x \in N$ be the disjunction of all the 
$\pi_p(x,y)$ whenever p has X as lefthand side.

We can write now the formula ${\overline{\pi}}_p(x,y)=$\\
$P_\alpha(x) \land P_\beta(y) \land \exists x_1 y_1 \ldots x_s y_s
[(x<x_1<y_1< \ldots < x_s<y_s<y)$\\
$\land 
(\psi_{v_0}(x,x_1)\land \psi_{v_1}(y_1,x_2)\land \ldots \land \psi_{v_s}(y_s,y))$\\
$ \land 
(M(x_1,y_1) \land \ldots \land M(x_s,y_s)) \land (\pi_{X_1}(x_1,y_1)\land \ldots \land
\pi_{X_s}(x_s,y_s)]$,

which restricts the pattern of the matching between $x$ and $y$ to
correspond to the matching of a production having the appropriate non terminal as left
hand side.

The formula $\psi_G$ is then:
$$\bigvee_{S \rightarrow u \in P}(\psi_u(min,max)) \lor [\forall x \forall y
(M(x,y) \rightarrow$$
$$ \bigvee_{p\in P}{\overline{\pi}}_p(x,y)) \land (M(min,max)
\land \pi_S(min,max))]$$

Since every production is uniquely determined by its pattern, this formula is
appropriate for our aim.

For the other direction we remark that the construction of the tree is intimately connected to the matching. 
Then the uniqueness of the matching implies the uniqueness of the derivation tree for each word, 
this gives us, by definition, the unambiguity of the language. {\bf Q.E.D.}

{\bf Note.} As the property ''\emph{a binary relation is a matching}'' can be expressed in first-order logic, 
we can construct a syntactic sublogic of $IMP_2$ which captures Unambiguous Context Free Languages. 
This can be done by the set of formulas $\phi \land \psi$, where $\phi$ defines a binary relation implicitely and 
$\psi$ test if this relation is a matching. We gave in this paper the semantic definition rather than the syntactic 
one because of the simplicity of this notion in this case.

\begin{cor}
$IMP_2$ is undecidable.
\end{cor}

This is a simple consequence of undecidability of unambiguity.

\section{Conclusion}

We reproved in this paper an algebraic characterization of Context Free Languages by means of Dyck languages, using a result of McNaughton and Papert \cite{MP} for the logical description of star free expressions and the Double Greibach Normal Form. We could get a cleaner proof by using the Double Quadratic Greibach Normal Form.

Unambiguity of Context Free languages is relevant for compiling theory because if a program has two different derivations we can have different results for the same input.

This motivates me to try to describe Unambiguous Context Free Languages by logical means. But the undecidability of Unambiguity compels us to use an undecidable logic, which is $IMP$. For a proof of its undecidability see \cite{Kol}.

The undecidability of $IMP$ is in the sense commonly understood.
 That is the set of $IMP$-formulas is {\it co-recursively enumerable complete}. But the undecidability of $IMP_2$ is in the sense that we can't decide if a given binary predicate, 
which is a matching, can be whether or not implicitly defined by a first-order formula.

The result of Eiter and {\it al} \cite{GGG} discouraged me to look for some more syntactic logic for all classes between $N.P.$ and regular sets.

The result makes the link between two undecidable problems, a logical one and a language theoretic one.

The question which naturally arises after our result is:

{\it Is there a logic for deterministic Context Free Languages?}


\begin{thebibliography}{99}
\bibitem{aut}AUTEBERT, Jean Michel.
\emph{Th\'eorie des langages et des automates}, Masson 1994.
\bibitem{AA}AUTEBERT, Jean Michel.
\emph{Personnal communication}, 1998.
\bibitem{ABB}AUTEBERT, Jean-Michel, BERSTEL, Jean, et BOASSON, Luc. \emph{Context-free languages and pushdown automata.}
 In : Handbook of formal languages. Springer Berlin Heidelberg, 1997. p. 111-174.
\bibitem{buc}B\"UCHI, J. Richard. \emph{Weak Second-Order Arithmetic and Finite Automata.}
 Mathematical Logic Quarterly, 1960, vol. 6, no 1-6, p. 66-92.
\bibitem{CS}SCHUTZENBERGER, M. P. \emph{THE ALGEBRAIC THEORY OF CONTEXT-FREE LANGUAGES* N. CHOMSKY.}
 Computer programming and formal systems, 1963, vol. 28, p. 118.
\bibitem{DD}DONER, John. \emph{Tree acceptors and some of their applications}. Journal of Computer and System Sciences, 1970, vol. 4, no 5, p. 406-451.
\bibitem{EF}EBBINGHAUS, Heinz-Dieter et FLUM, J\"org. \emph{Finite model theory.} Springer Science \& Business Media, 2005.
\bibitem{GGG}EITER, Thomas, GOTTLOB, Georg, et GUREVICH, Yuri. \emph{Existential second-order logic over strings.} 
Journal of the ACM (JACM), 2000, vol. 47, no 1, p. 77-131.
\bibitem{elg}ELGOT, Calvin C. \emph{Decision problems of finite automata design and related arithmetics.} Transactions of the American Mathematical Society, 1961, vol. 98, no 1, p. 21-51.
\bibitem{fagin}FAGIN, Ronald.
\emph{Generalized first-order spectra and polynomial time  recognizable sets}, in RM Karp editor Complexity of computation, SIAM-AMS Proceedings 1974.
\bibitem{GS}GÉCSEG, Ferenc et STEINBY, Magnus. \emph{Tree languages.} In : Handbook of formal languages. Springer Berlin Heidelberg, 1997. p. 1-68.
\bibitem{HY1}HACHA\"ICHI, Yassine. \emph{A descriptive complexity approach to the linear hierarchy.} Theoretical computer science, 2003, vol. 304, no 1, p. 421-429.
\bibitem{HY2}HACHA\"ICHI, Yassine. \emph{Fragments of monadic second-order logics over word structures.} Electronic Notes in Theoretical Computer Science, 2005, vol. 123, p. 111-123.
\bibitem{Ha}HARRISON, Michael A. \emph{Introduction to formal language theory.} Addison-Wesley Longman Publishing Co., Inc., 1978.
\bibitem{GH}HOTZ, Guenter. \emph{Normal-form transformations of context-free grammars}. Acta Cybernetica, 1980, vol. 4, p. 65-84.
\bibitem{Kol}KOLAITIS, Phokion G. \emph{Implicit definability on finite structures and unambiguous computations}. In : Logic in Computer Science, 1990. LICS'90, Proceedings., Fifth Annual IEEE Symposium on e. IEEE, 1990. p. 168-180.
\bibitem{LA} LACROIX, Zo\'e.
\emph{Bases de donn\'ees des relations implicites aux relations contraintes}, Ph.D. Universit\'e de Paris Sud 1996.
\bibitem{LST}LAUTEMANN, Clemens, SCHWENTICK, Thomas, et TH\'ERIEN, Denis. 
\emph{Logics for context-free languages.} In : Computer science logic. Springer Berlin Heidelberg, 1994. p. 205-216.
\bibitem{MP}MCNAUGHTON, Robert et PAPERT, Seymour A. \emph{Counter-Free Automata} (MIT research monograph no. 65). The MIT Press, 1971.
\bibitem{MW}MEZEI, Jorge et WRIGHT, Jesse B. \emph{Algebraic automata and context-free sets.}
 Information and control, 1967, vol. 11, no 1, p. 3-29.
\bibitem{PAP} PAPADIMITRIOU, Christos H. \emph{Computational complexity.} John Wiley and Sons Ltd., 2003.
\bibitem{Pi}PIN, Jean-Eric. \emph{Logic, semigroups and automata on words}
. Annals of Mathematics and Artificial Intelligence, 1996, vol. 16, no 1, p. 343-384.
\bibitem{Str}STRAUBING, Howard. \emph{Finite automata, formal logic, and circuit complexity.} Springer Science \& Business Media, 2012.
\bibitem{TW}THATCHER, James W. et WRIGHT, Jesse B.. \emph{Generalized finite automata theory with an application to a decision problem of second-order logic.} Mathematical systems theory, 1968, vol. 2, no 1, p. 57-81.
\bibitem{Th}THOMAS, Wolfgang. \emph{Languages, automata, and logic,} Handbook of formal languages, vol. 3: beyond words. 1997.
\end{thebibliography}
\end{document}